\documentclass[aps,twocolumn,showpacs,pra,amsmath,amssymb,floatfix,superscriptaddress]{revtex4-1}
\usepackage{amsmath,amssymb,graphicx,color}
\usepackage{bm}
\usepackage{bbm}
\usepackage[caption=false]{subfig}
\usepackage[usenames,dvipsnames]{xcolor}
\usepackage[normalem]{ulem}
\usepackage{soul}
\usepackage[colorlinks=true,citecolor=Cerulean,linkcolor=RubineRed,urlcolor=Cerulean]{hyperref}
\usepackage{soul,xcolor}

\usepackage{mathptmx} 

\begin{document}
\title{Floquet eigenspectra of a nonlinear two-mode system under periodic driving:\\ the emergence of ``ring" structures}

\author{Guitao Lyu} \email{guitao@zju.edu.cn}
\affiliation{Department of Physics and Zhejiang Institute of Modern Physics, Zhejiang University, Hangzhou, Zhejiang 310027, China}
\author{Lih-King Lim}
\affiliation{Department of Physics and Zhejiang Institute of Modern Physics, Zhejiang University, Hangzhou, Zhejiang 310027, China}
\author{Gentaro Watanabe}\email{gentaro@zju.edu.cn}
\affiliation{Department of Physics and Zhejiang Institute of Modern Physics, Zhejiang University, Hangzhou, Zhejiang 310027, China}
\affiliation{Zhejiang Province Key Laboratory of Quantum Technology and Device, Zhejiang University, Hangzhou, Zhejiang 310027, China}

\begin{abstract}
We study Floquet eigenspectra of a nonlinear two-mode system under a periodic driving of the off-diagonal coupling. By solving the Gross-Pitaevskii equation numerically, we obtain triangular and loop structures near the crossings of different Floquet branches. At lower driving frequencies, we find ``ring" and ``double-ring" structures which are distinct from the well-known loop structure. The mechanism of the emergence of these structures is discussed and the parameter windows of their existence are obtained analytically. In addition, we study the evolution of the system under the driving with an adiabatic sweep and find there are some dynamically unstable states in the Floquet eigenspectra which break the quantum adiabaticity.
\end{abstract}

\maketitle

\section{Introduction}
The two-mode system is a paradigm to study many fundamental quantum phenomena, including many facets of Landau-Zener physics \cite{Peter Hanggi, Tunneling between two BEC, two bands of BEC, soliton NLZ, LZT in wga, observation of LZ in wg} and Josephson effects \cite{JE two bec, Josephson Effects, Leggett01, Observation of Josephson}. With the experimental progress of cold atoms, a Bose-Einstein condensate (BEC) loaded in a double-well potential introduces an additional ingredient, namely, nonlinear effects due to interactions, described by a two-mode Gross-Pitaevskii equation (GPE). Analogous nonlinear wave equations with mode coupling are also used to describe a class of photonic lattices \cite{Garanovich2012, wg with Kerr, Nonlinear LZT wga} although the origin of nonlinearity is distinct, i.e., the nonlinear Kerr effect. Combining the basic Landau-Zener process with nonlinearity gives rise to effects deemed counterintuitive for quantum linear systems, such as the occurrence of a loop energy spectrum and the resulting breakdown of quantum adiabaticity \cite{double-well nphys, Biao Wu NLZ, Biao Wu and Jie liu, generalized LZ formula,Bomantara17}.

Recently, in advancing the study of quantum dynamics, time-periodically driven quantum systems have received renewed attention shedding new insights on out-of-equilibrium quantum matter \cite{Eckardt17,Oka19}. A first step towards the understanding of a driven interacting bosonic system is the study of level transitions in the presence of a self-consistent mean-field interaction. On the one hand, there have been numerous theoretical studies of nonlinear Landau-Zener \cite{Biao Wu NLZ, Biao Wu and Jie liu} or interacting two-mode Bose systems with periodic modulations in either the level spacing \cite{Holthaus01a, Holthaus01b, Kelin Gao chaos 2001, Kohler03, Photon-Assisted Tunneling, Floquet Quasienergies, Jie Liu self-trapping, Xie, Xie2, Yosuke Kayanuma 2008, C. Weiss chaos 2008, Jiangbin Gong odd even 2009} or an off-diagonal coupling \cite{Salmond02, Jiangbin Gong, Jiangbin Gong njp, Watanabe10, equilibrium state and chaos}, or both  \cite{F. Kh. Abdullaev 2000, Haroutyunyan04, Watanabe12}. Phenomena such as the coherent destruction of tunneling (originally studied in Refs.~\cite{Dunlap86, Grossmann91, Holthaus92} for different setups) realizing a dynamical localization \cite{Kierig08, Yosuke Kayanuma 2008, Jiangbin Gong odd even 2009, Watanabe12}, macroscopic self-trapping \cite{Tunneling between two BEC, F. Kh. Abdullaev 2000, Holthaus01a, Holthaus01b, Jie Liu self-trapping, Xie2, Xie}, assisted higher-order co-tunneling \cite{Watanabe10, Watanabe12}, as well as the emergence of Hamiltonian chaos \cite{Kelin Gao chaos 2001, Salmond02, C. Weiss chaos 2008, Jiangbin Gong njp, equilibrium state and chaos} have been uncovered. The last one makes an intriguing connection to studies of dynamical systems with chaos \cite{equilibrium state and chaos}. On the other hand, given the favorable experimental timescale in both the manipulation and observation, nonlinear quantum dynamics becomes relevant in periodically driven BEC systems, which have fundamental applications in quantum metrology \cite{Oberthaler08}. 

In the present paper, we study a nonlinear two-mode system under a time-periodic driving of the off-diagonal coupling (so-called off-diagonal driving) with Floquet analysis. The validity of Floquet analysis for such a nonlinear system has been confirmed in Refs.~\cite{Holthaus01a, Holthaus01b} with the help of the Poincar\'e-Birkhoff theorem. The appeal of the Floquet approach \cite{Peter Hanggi, Beyond Floquet theorem, Floquet engineering, Holthaus16} is the ability to disentangle the effect of band couplings, which typically involves more than two Floquet bands, from the nonlinear effect due to the self-consistent interactions. We numerically solve the nonlinear Schr\"odinger equation (as realized by GPE for a two-mode BEC) focusing on a few regimes (from high- to low-frequency driving) where complex eigenspectrum structures can emerge. While the Floquet analysis has been applied to the nonlinear two-mode system by many authors \cite{Holthaus01a, Holthaus01b, Salmond02, Xie2, Floquet Quasienergies, Strzys08, Xie, equilibrium state and chaos}, only a few works have studied the Floquet quasienergy spectrum in depth. Even in the latter, they focus only on the high-frequency regime \cite{Floquet Quasienergies}, or employ an effective Hamiltonian description, which is valid only for high-frequency driving, without a Floquet analysis \cite{Jiangbin Gong, Jiangbin Gong njp}. Here, our work is no longer restricted to a particular frequency range, provided that the solution assumes a Floquet form. Focusing on the topology of the eigenspectrum, we find ``ring" and ``multiple-ring" structures caused by a combination of nonlinearity and the coupling between Floquet branches. In addition, we briefly study the time evolution of the system under the periodic driving with an adiabatic sweep of the level spacing. We find signatures of dynamical instability in the quasienergy bands where ``adiabaticity breakdown" is observed.

This paper is organized as follows. In Sec. II, we introduce the nonlinear two-mode model under a periodic driving of the off-diagonal coupling between the two modes. In Sec. III, we present the triangular and ``ring" structures in Floquet eigenspectra for various parameters. We clarify the mechanism of the emergence of the ``ring" structure and give parameter windows of their existence. In Sec. IV, the adiabatic evolution of the nonlinear two-mode system under the off-diagonal driving are discussed. In Sec. V, we summarize our work and discuss the feasibility of our predictions in experiments. Throughout the paper, we set $\hbar = 1$.

\section{Model}
We consider a BEC trapped in a double-well potential with mode $a$ and $b$ in the Gross-Pitaevskii mean-field description, and a time-periodic modulation of the off-diagonal coupling in the form of $\delta v\, e^{\pm i\omega t}$, where $\delta v$ is the driving strength and $\omega$ is the driving frequency. This form of the driving can be realized, e.g., by a tilted double-well potential in a rotating frame \cite{note:drive,note:accellattice}. The system can be described by the following Hamiltonian \cite{Biao Wu NLZ}: 
\begin{align}
H(t)=\frac{1}{2}
\begin{bmatrix}
\gamma+g(\left| b \right|^2-\left| a \right|^2) & v+\delta v\, e^{i\omega t}
\\ v +\delta v\, e^{-i\omega t} & -\gamma - g (\left| b \right|^2-\left| a \right|^2)
\end{bmatrix}, 
\label{eq:hh1} 
\end{align}
where $a=a(t)$ and $b=b(t)$ are the amplitudes of the condensate wave function of modes $a$ and $b$, respectively, $\gamma$ is the level spacing between the two wells, and $g$ is the two-body interaction strength of atoms in each well. Note that the energy difference of the two modes also depends on the population difference $ \left| b(t) \right|^2-\left| a(t) \right|^2 $ between modes $a$ and $b$, coming from the mean-field interactions. $v$ is the coupling strength of the two modes depending on the barrier height between the two wells. Throughout this paper, we set the coupling strength $v=1$ and take $v$ as the unit of energy and $1/v$ as the unit of time. In Sec.~\ref{sec.3a}, we also consider an off-diagonal driving with the form of $\delta v\, \cos(\omega t)$, which represents a periodic modulation of the barrier height of the double-well potential. The time evolution of the system is described by the time-dependent Gross-Pitaevskii equation:
\begin{align}
i \partial_t \psi(t) = H(t)\, \psi(t),
\label{eq:hh2}
\end{align}
where $\psi(t)\equiv (a(t),\ b(t))^ \mathrm{T} $ is the condensate wave function, which is to be determined self-consistently (see below). In the absence of the interaction terms, the Hamiltonian is time periodic with period $T=2\pi /\omega$ of the driving:
\begin{align}
H(t)=H(t+T).
\label{eq:hh3}
\end{align}
According to the Floquet theorem, we take the solutions of the GPE in the following form:
\begin{align}
\psi(t)=e^{-i\epsilon t}\widetilde{\psi}(t).
\label{eq:hh4}
\end{align}
Here $\epsilon$ is the quasienergy and $\widetilde{\psi}(t) \equiv (\widetilde{a}(t),\ \widetilde{b}(t))^ \mathrm{T} $ is a time periodic function whose period is the same as that of the driving:
\begin{align}
\widetilde{\psi}(t)=\widetilde{\psi}(t+T).
\label{eq:hh5}
\end{align}
Because of the time-periodicity, we expand
$\widetilde{\psi}(t)$ in a Fourier series:
\begin{align}
\widetilde{a}(t)=\sum\limits_{n=-{\infty}}^{+\infty} c_{n}\, e^{i n \omega t},
\label{eq:hh6}\\
\widetilde{b}(t)=\sum\limits_{n=-\infty}^{+\infty} d_{n}\, e^{i n \omega t},
\label{eq:hh7}
\end{align}
where $n$ is an integer, and recast the original GPE (\ref{eq:hh2}) into an eigenvalue problem, albeit nonlinear, for the coefficients $\{c_n,\, d_n\}$. They are to be solved iteratively numerically, in a self-consistent manner. In actual numerical calculations, the summation with respect to $n$ is truncated at a cutoff value $\pm n_{\rm max}$ when convergence is achieved. From Eqs.~(\ref{eq:hh4}), (\ref{eq:hh6}), and (\ref{eq:hh7}), we see that the quasienergy $\epsilon$ has a periodic structure with period $\omega$ as a result of the Floquet theorem. Finally, since the total number of particles is conserved, the coefficients $c_{n}$ and $d_{n}$ should satisfy the additional constraint:
\begin{align}
\sum\limits_{n} \left| c_{n} \right|^2 +  \sum\limits_{n} \left| d_{n} \right|^2 =1\,.
\label{eq:hh8}
\end{align}

\section{Floquet eigenspectra} 

\subsection{Periodicity and triangular structures \label{sec.3a}}
\begin{figure*}[th!]
	\centering \includegraphics[scale=0.86]{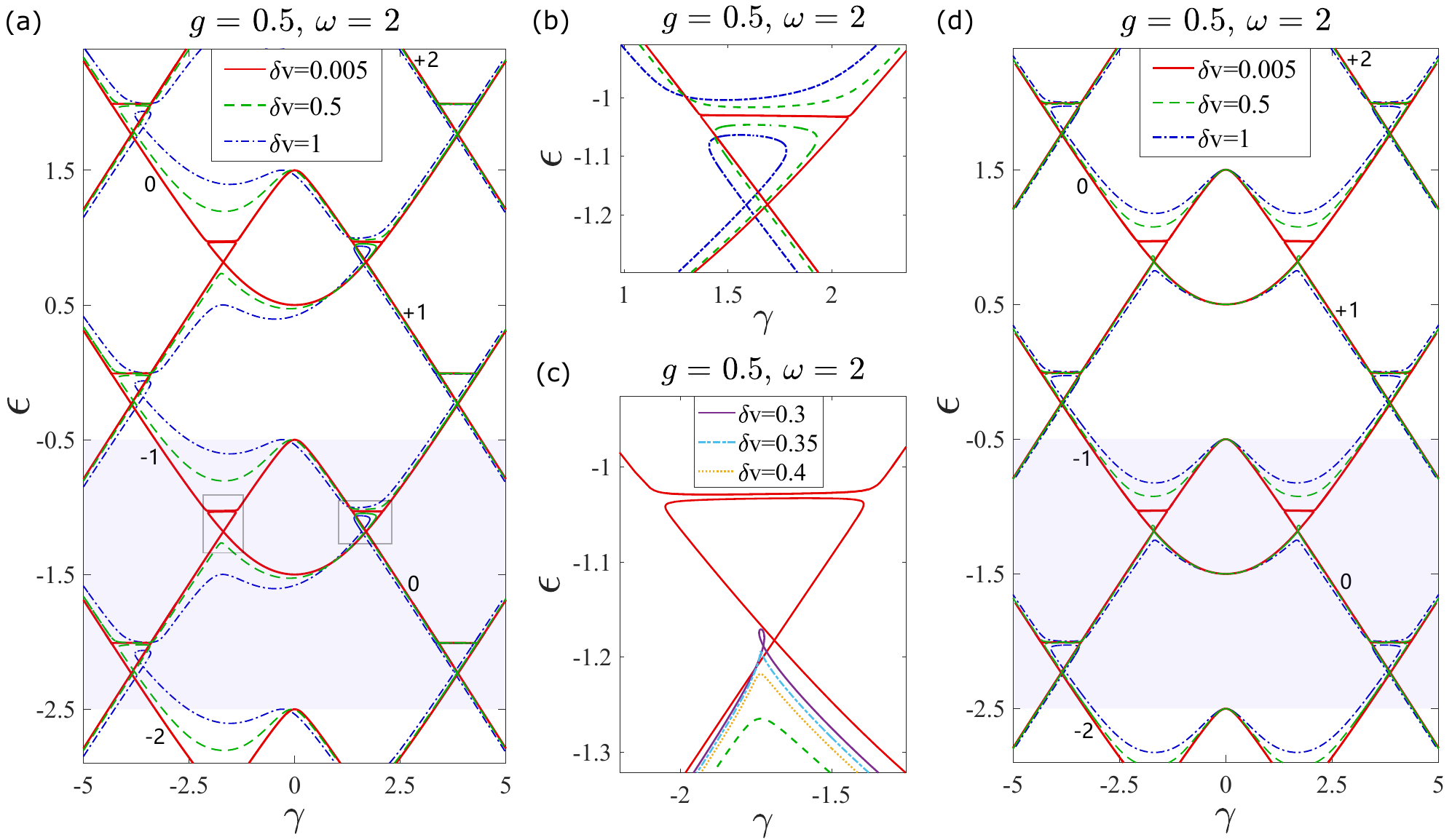} \caption{Floquet eigenspectra as functions of the level spacing $\gamma$ at a fixed frequency $\omega=2$ and interaction strength $g=0.5$, for various driving strength: $\delta v= 0.005$ (red solid line), $0.5$ (green dashed line), and $1$ (blue dashed dotted line). (a) Periodic eigenspectra for the driving in the form of $\delta v\, e^{\pm i\omega t}$. The shaded region is one ``Brillouin zone". (b) A magnified view of the right gray box area in (a), showing the triangular structure near the crossing. (c) A magnification of the left gray box area in (a) with additional values of the driving strength $\delta v = 0.3$, $0.35$, and $0.4$. (d) Periodic eigenspectra for the driving in the form of $\delta v\, \cos(\omega t)$.}
	\label{fig:1}
\end{figure*}

In a two-level system without the off-diagonal coupling and driving, $ v=\delta v=0 $, the energy spectrum as a function of the level spacing $\gamma$ shows a crossing at $\gamma=0$. Once the coupling $v$ between the two modes is set to nonzero, a gap opens in the crossing region of $\gamma$, so that the crossing turns into an avoided crossing. With regard to a double-well BEC system, when the atom-atom interaction strength $g$ is greater than the coupling strength $v$ between the two modes of the BEC, a loop appears in the avoided crossing region due to nonlinear effects, which has been discussed in Refs.~\cite{Biao Wu NLZ, Biao Wu and Jie liu}. In our model, besides the intrinsic nonlinearity, the time-periodic off-diagonal driving couples different Floquet branches. In this subsection, we focus on a frequency range larger than the characteristic energy gap of the two-level system, given by $v$ at $\gamma=0$; in particular, we show results for $\omega=2$ as an example. We numerically solve the time-dependent GPE (\ref{eq:hh2}) and get the Floquet eigenspectra $\epsilon(\gamma)$ shown in Fig.~\ref{fig:1}(a) for various values of the driving strength $\delta v$ with a fixed interaction strength $g=0.5$.

Firstly, focusing on the case of weak driving strength $\delta v =0.005$ (red line), we can clearly see the periodic structure (the shaded region shows a ``Brillouin zone") in the quasienergy space which can be understood by translating the two branches in the case without driving by $\pm n\omega$ ($n \in \mathbb{N}$) in the vertical direction. In the following discussion, we always treat these two branches as reference branches since they do not move under the change of the driving frequency $\omega$ unlike the other branches. In fact, with the two reference branches, the dominant Fourier components of modes $a$ and $b$ are $c_{0}$ and $d_{0}$, and the ratio between $\left| c_{0} \right|^2$ and  $\left| d_{0} \right|^2$ changes with the level spacing $\gamma$. The populations of modes $a$ and $b$ are equal (i.e., $|a|^2=|b|^2=1/2$) at the top (bottom) of the lower (upper) reference branches at $\gamma=0$ in the limit of $\delta v =0$. Moreover, for any $\delta v$, the relative phase between the two modes at the lower (upper) reference branch is $\pi$ ($0$) when $t=mT$ $(m \in \mathbb{Z})$ \cite{note:oscillation}. (See also Ref.~\cite{Biao Wu and Jie liu} for discussions on the case without driving.)

The other Floquet branches, generated by shifting up (down) the reference branches by $ n\omega$, have dominant Fourier components $c_{n}$ and $d_{n}$ ($c_{-n}$ and $d_{-n}$), and share the same population ratio and relative phase between the two modes with the reference branches. Hence, for convenience, we label the reference branches by $0$, and the other branches by $\pm n$ [see, e.g., the labeled number in Fig.~\ref{fig:1}(a)]. Note that the labeling of the Floquet branch is based on the case in the limit of $\delta v=0$: each Floquet branch labeled by an integer $\pm n$ is a continuous function of $\gamma$ and extends over a (semi)-infinite range of the quasienergy when $\delta v=0$. In the case of $\delta v \ne 0$, the Floquet branches are separated into different quasienergy bands extending within a finite range of the quasienergy, but we still maintain this labeling for the Floquet branches [see, e.g., Fig.~\ref{fig:1}(a)].

Next, we can see the ``triangular" structures in the eigenspectrum at $\gamma\approx \pm 2$ [see Figs. 1(b) and 1(c) for a magnified view of the right and left gray boxes area in Fig. 1(a), respectively]. Figure~\ref{fig:1}(c) demonstrates the evolution of the triangular structure by changing $\delta v$. As $\delta v$ increases and $g$ fixed, the triangular structure gradually turns into a loop and a cusp for $\delta v \approx g$. For $\delta v \gtrsim g$, the quasienergy spectrum evolves into an avoided crossing with an energy gap roughly proportional to the driving strength $\delta v$ as nonlinear effect becomes insignificant.

Actually, the triangular and the loop structures at $\gamma\approx \pm 2$ are similar to the loop structure found in Ref.~\cite{Biao Wu NLZ}, except that the former is due to the competition between $g$ and $\delta v$ while the latter is due to the competition between $g$ and $v$. Indeed, at $\gamma\approx \pm 2 $ (and $\omega=2$), the Floquet matrix of $H(t) - i \partial_t$ with small $g$ can be approximated to an effective two-level nonlinear problem dressed by a coupling $\delta v$ between two neighboring Floquet branches, which takes a similar form of the $2\times 2$ Hamiltonian matrix studied in \cite{Biao Wu NLZ}. In addition, there are other triangular structures at $\gamma\approx \pm 4$, $\pm 6$, $\cdots$ in Fig.~\ref{fig:1}(a), which come from the nonlinear crossings between the two Floquet branches with absolute difference of their indices being $2$, $3$, $\cdots$. However, in these cases, there is no direct coupling by $\delta v$ between these non-neighboring Floquet branches, and the coupling becomes higher order of $\delta v$ in the effective two-level nonlinear problem, so that the nonlinear effects are more apparent. As a result, these triangular structures still remain even for $\delta v \gtrsim g$ [see the triangular structures at $\gamma\approx \pm 4 $ in Fig.~\ref{fig:1}(a)].

Lastly, one notices that the Floquet eigenspectra shown in Fig.~\ref{fig:1}(a) are not symmetric with respect to $\gamma=0$. In fact, the asymmetry comes from the fact that the phase factors $e^{\pm i\omega t}$  in the off-diagonal terms of the Hamiltonian (\ref{eq:hh1}) are not the same. If we consider the off-diagonal driving in the form of $\delta v\, \cos(\omega t)$, the eigenspectra will be symmetric with respect to $\gamma=0$ as shown in Fig.~\ref{fig:1}(d). In this case, the GPE (\ref{eq:hh2}) is invariant under the combined operation of replacing  $\gamma$ by $-\gamma$ and interchanging the modes $a$ and $b$. In other words, reversing the sign of the level spacing $\gamma$ as well as interchanging the two modes of the BEC, the system does not change physically. On the other hand, in the case of $\delta v\, e^{\pm i\omega t}$, the GPE is not invariant under the above operation due to the difference in the phase factors between the off-diagonal driving terms. Except for this asymmetry, there is no qualitative difference in the result between two forms of the driving. Hence, for simplicity, we avoid using large driving strength $\delta v$ in the following subsections in order to suppress the influence of the asymmetry effect.

\subsection{``Ring" structures \label{sec.3b}} 
\begin{figure}[htb!]
  \centering \includegraphics[scale=0.7]{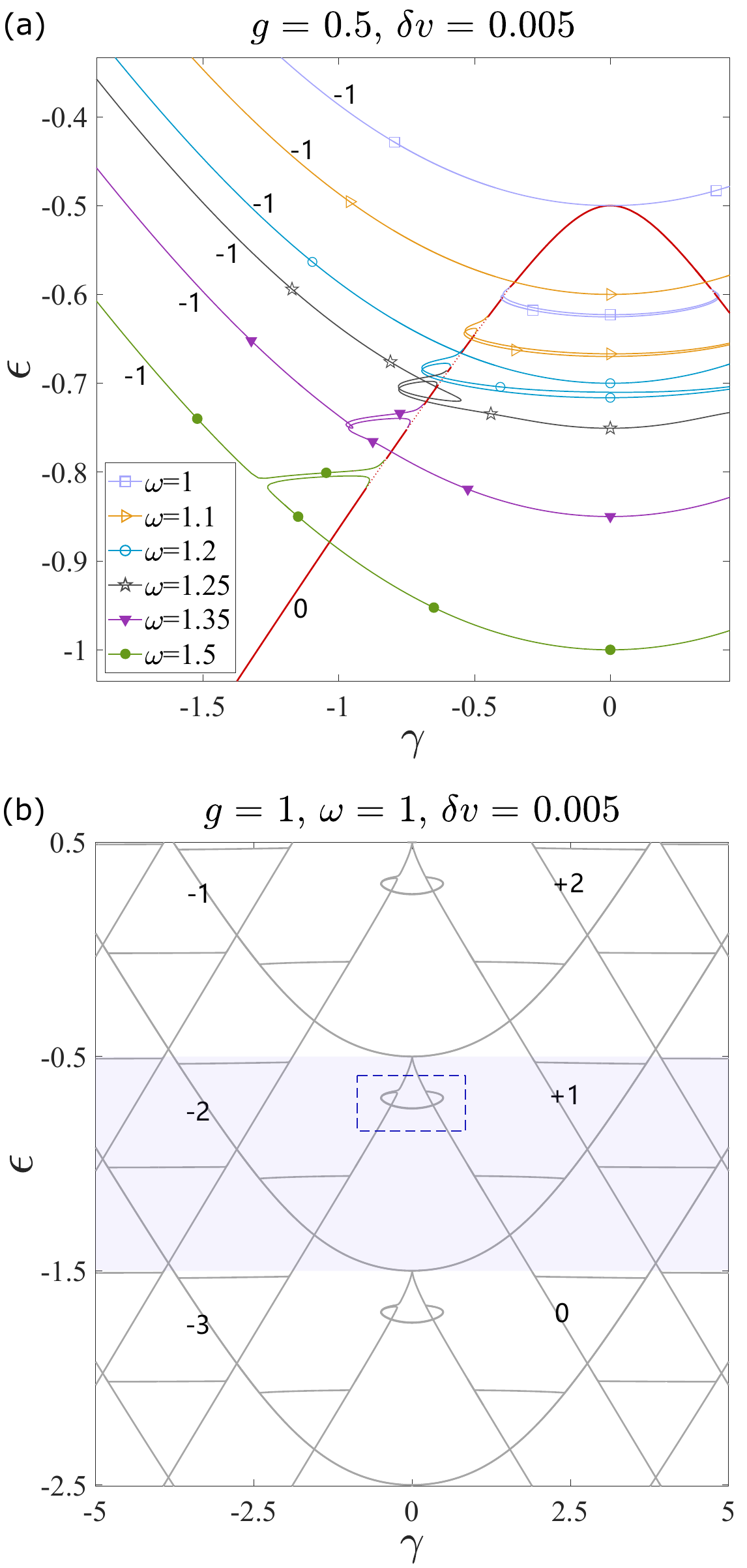} \caption{Emergence of the ``ring" structure near resonance. (a) Floquet spectra for various values of $1.5 \ge \omega \ge 1$ at a fixed $g=0.5$. The ring part appears when $\omega \lesssim 1.2$. The red line is the lower $0$ branch, which does not move by changing $\omega$. (b) Floquet spectrum for $\omega=1$ and a larger value of $g$ at $g=1$. The shaded region is one ``Brillouin zone", and the dashed box shows the ``ring" structure. Here we keep the driving strength suppressed at $\delta v=0.005$.
  }
	\label{fig:2}
\end{figure}

In this subsection, we turn to the on-resonance case where the driving frequency is equal to the off-diagonal coupling strength: $\omega = v$ $(= 1)$. In the cases around the on-resonance condition, we find a ``ring''-like structure shown, e.g., in the dashed box area in Fig.~\ref{fig:2}(b), which is distinct from the well-known loop structure (e.g., Refs. \cite{Biao Wu NLZ, Biao Wu and Jie liu, B. Wu OLBEC, Diakonov, Mueller, Machholm, G. W. rev}); hereafter we call this structure a ``ring".

To understand how the ring structure emerges, we follow the evolution of the spectrum as the driving frequency approaches the on-resonance condition from above as shown in Fig.~\ref{fig:2}(a). In this figure, the red line is the lower $0$ branch, which does not move by changing $\omega$. We can see that as $\omega$ decreases to $1.25$, the triangular structure slowly shrinks, and at a critical frequency value ($\omega\approx 1.2$, in blue) the upper $-1$ branch splits into two parts, giving rise to the ring part \cite{note:ring}. A further decrease in the frequency separates the ring part from the shifting $-1$ branch. Near the on-resonance condition, while the upper $-1$ branch touches the lower $0$ branch \cite{note:gap}, the ring structure still remains.

Taking a larger interaction strength around $g=1$ with $\omega$ fixed at $\omega=1$ as in Fig.~\ref{fig:2}(b), the ring part becomes more distinct, and the inverted parabola [i.e., the lower $0$ branch in Fig.~\ref{fig:2}(a)] becomes a cusp \cite{note:cusp}. In fact, the interaction strength $g=v$ $(=1)$ is a critical value for the emergence of the loop structure around $\gamma=0$ in the case without driving (i.e., $\delta v=0$), and the spectrum shows a cusp at $\gamma=0$ in this case. Therefore, we treat $g=1$ as a reference point which separates the regions of weak $(g < 1)$ and strong $(g > 1)$ nonlinearity. (The triangular structures observed in the regions away from $\gamma=0$ are the ones discussed previously.)

\begin{figure*}[th!] 
    \centering \includegraphics[scale=0.6]{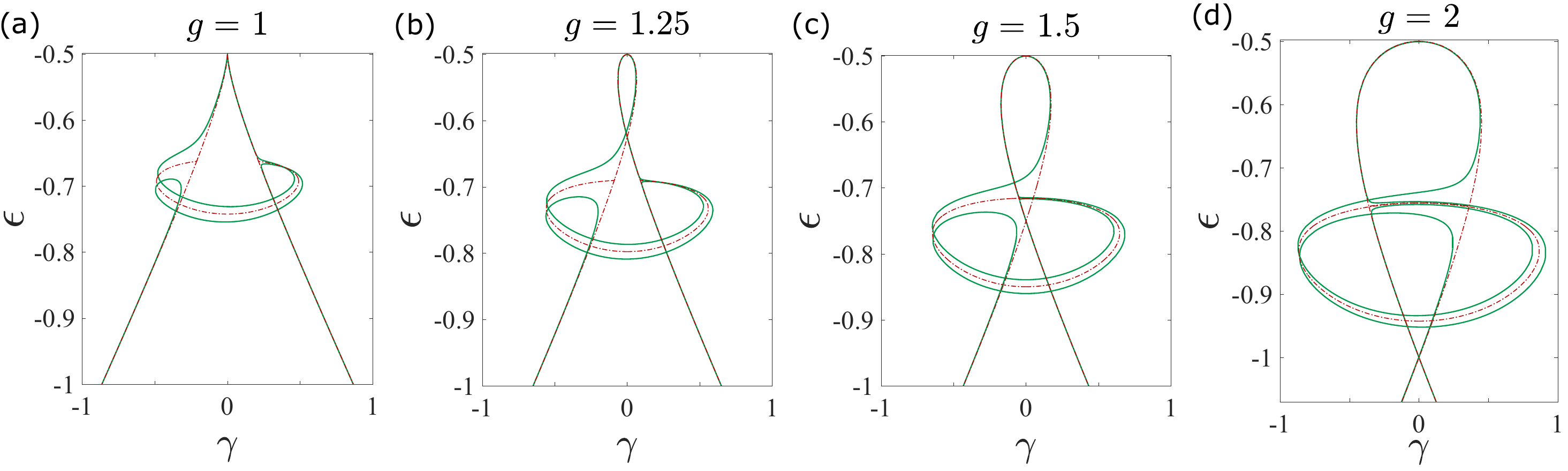} \caption{The ring merges with the loop with increasing $g=1$, $1.25$, $1.5$, and $2$, as shown in (a)--(d), respectively. Here we keep $\omega=1$ and take $\delta v=0.05$ (in green solid curves), which is $10$ times larger than the value used in Fig.~\ref{fig:2} to clearly show the splitting due to nonzero $\delta v$. For demonstrating the split of the degeneracy of the ring part by nonzero $\delta v$, we also show the ring structures obtained by setting $\delta v=0$ (in red dash-dotted curves) but maintaining the periodicity of $\widetilde{a}(t)$ and $\widetilde{b}(t)$.}  
	\label{fig:3}
\end{figure*}

As for the dependence on the nonlinear effect of the ring structure, there are two consequences: (1) by increasing, or (2) by decreasing the interaction strength from $g=1$. In the first case, it basically mixes the ring and the cusp structures further; see Fig.~\ref{fig:3}. Specifically, as the cusp turns into a loop structure, the ring gets bigger and merges with the loop. As shown in Fig.~\ref{fig:3}(d), for sufficiently large interaction strength, the intertwined spectrum takes a shape like the ``G clef" symbol. In combination with the asymmetry effect of the off-diagonal driving form $e^{\pm i\omega t}$, a distorted ``ring-loop-mixing" structure emerges [see, e.g., the green solid curves in Fig.~\ref{fig:3}(d)].

For comparison, we also plot, in the same figures, the spectra obtained by setting $\delta v=0$ but maintaining the periodic Floquet form of $\widetilde{a}(t)$ and $\widetilde{b}(t)$ given by Eqs.~(\ref{eq:hh6}) and (\ref{eq:hh7}) [red dotted curves in Fig.~\ref{fig:3}]. It is a good approximation when $\delta v$ is negligible compared with the other parameters \cite{G. W. fermi swallow-tail}. We see that the fine features such as two ``rings" (green solid curve) seen in the presence of a nonzero $\delta v$ are lost and the Floquet spectral curves become degenerate.

\begin{figure}[tb!]
	\centering \includegraphics[scale=0.5]{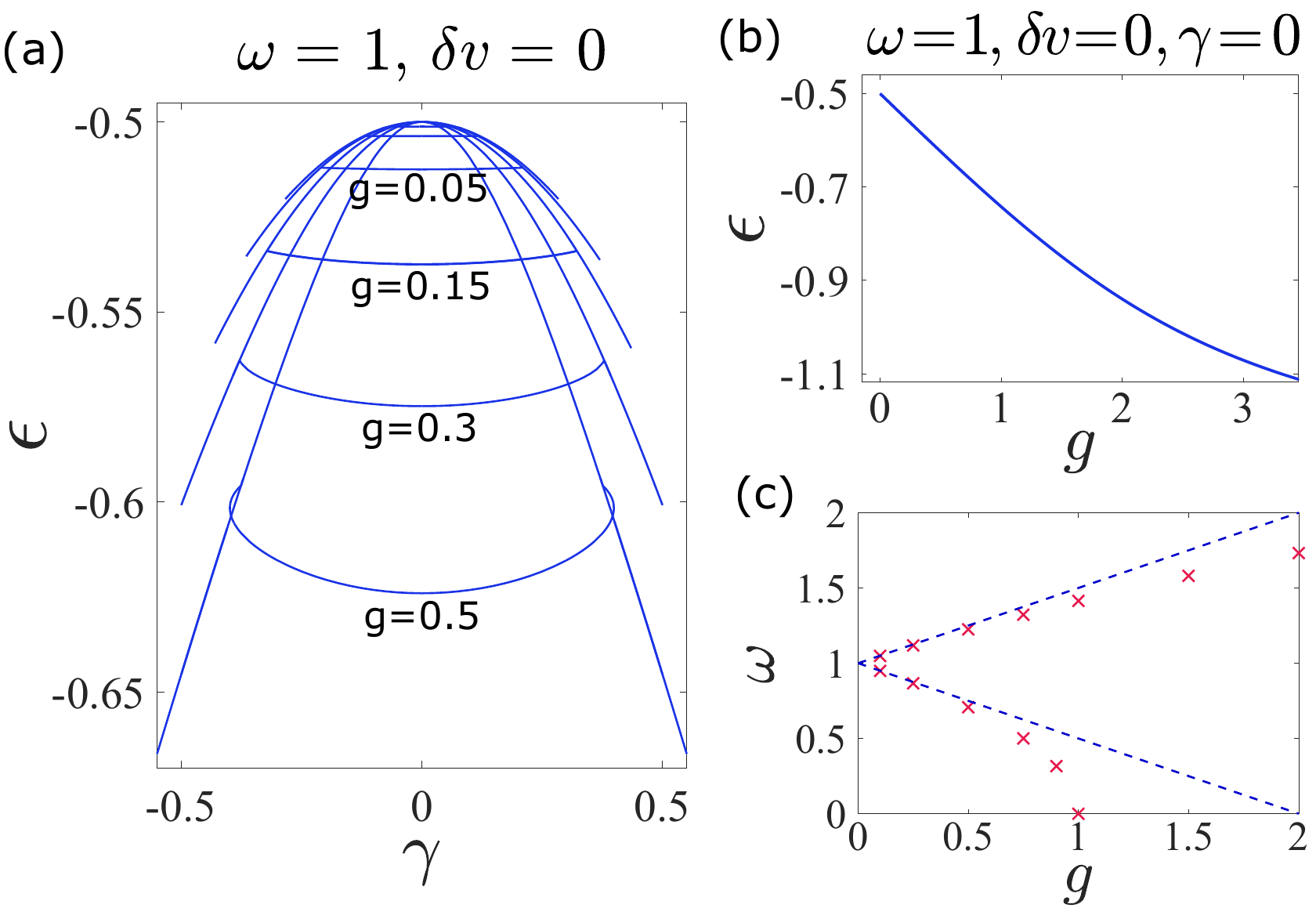} \caption{Disappearance of the ring structure and the parameter window of its existence. Here we employ the simplified analysis with $\delta v=0$. (a) With decreasing $g$, the ring shrinks to the top of the lower $0$ branch and disappears when $g=0$. (b) The quasienergy of the ring part at $\gamma=0$ as a function of $g$. (c) Frequency window of the existence of the ring as a function of the interaction strength $g$. The upper (lower) dashed line shows the upper (lower) critical frequency given by Eq.~(\ref{eq:window}), and the red crosses show the numerical results.}
	\label{fig:4}
\end{figure}

In the case of decreasing interaction strength from $g=1$, we expect the ring part may disappear at a critical value of $g$. Indeed, as shown in Fig.~\ref{fig:4}(a), the cusp gets rounded to a smooth inverted parabola, and the ring part gradually shrinks to the top of the lower $0$ branch. To follow the evolution of the ring position in the quasienergy space, we plot the quasienergy $\epsilon$ on the ring at $\gamma=0$ as a function of the interaction strength $g$ in Fig.~\ref{fig:4}(b). We see that, as $g$ decreases to $0$, the ring approaches the top ($\epsilon=-0.5$) of the lower $0$ branch [see also Fig.~\ref{fig:4}(a)]. In other words, as the nonlinear effects disappear, so does the ring part; the presence of the nonlinearity is thus a necessary condition for the emergence of the ring structures.

To gain a better understanding on the emergence of the ring structure, we perform a simple analysis on the structure of the wave function $\psi(t)$. Focusing on the Floquet eigenstates of the ring parts close to the emergence point [i.e., the top point of the lower $0$ branch in Fig.~\ref{fig:4}(a)], we find that the solution of the ring part has four dominant Fourier components, $c_{-1}$, $c_{0}$, $d_{-1}$, and $d_{0}$, and they satisfy the following relations at $\gamma =0$:
\begin{align}
c_{-1}=d_{-1} \qquad \mathrm{and} \qquad c_{0}=-d_{0}.
\label{eq:hh9}
\end{align}

As we mentioned in Sec.~\ref{sec.3a}, each Floquet branch should have only one dominant Fourier component for each mode. However, there are two dominant Fourier components for each mode in the ring parts, which is caused by the hybridizations of Floquet branches. It is noted that, in the limit of $\delta v=0$, the populations of modes $a$ and $b$ are equal at the bottom (top) of upper $-1$ (lower $0$) branch at $\gamma=0$. In addition, the relative phase between modes $a$ and $b$ of the upper $-1$ branch is $0$ and that of the lower $0$ branch is $\pi$. As a result, $c_{-1}=d_{-1}$ and $c_{0}=-d_{0}$, respectively. Hence the ring part at $\gamma=0$ satisfies Eq.~(\ref{eq:hh9}). 

Now we make the following approximation on the ring part at $\gamma =0$: Take Eq.~(\ref{eq:hh9}), and neglect the other components of modes $a$ and $b$. Furthermore, we set $\delta v=0$ and keep the periodicity of the quasienergy. Under these simplifications, we insert Eqs.~(\ref{eq:hh6}) and (\ref{eq:hh7}) into Eq.~(\ref{eq:hh2}) to get four equations for $c_{-1}$, $c_{0}$, $d_{-1}$, and $d_{0}$; taking the symmetry into account, the number of equations is halved. Finally, we get the following two relations: 
\begin{align}
 g\left| c_{-1} \right|^2  + \epsilon + \frac{v}{2} &=0,
\label{eq:hh10}\\
g\left| c_{0} \right|^2 + \omega + \epsilon  - \frac{v}{2} &=0,
\label{eq:hh11}
\end{align}
which give the parameter window of the emergence of ring structure. From Eqs.~(\ref{eq:hh8}) and (\ref{eq:hh9}), one gets
\begin{align}
 \left| c_{-1} \right|^2 + \left| c_{0} \right|^2 =\frac{1}{2}\,.
 \label{eq:hh12}
\end{align}
The ratio between $\left| c_{-1} \right|^2$ and $\left| c_{0} \right|^2$ reflects the extent of the hybridization between the upper $-1$ branch and the lower $0$ branch. The critical condition of disappearance of the ring structure is that either $\left| c_{-1} \right|^2$ or $\left| c_{0} \right|^2$ vanishes. $\left| c_{-1} \right|^2$ decreases to $0$ and $\left| c_{0} \right|^2$ increases to $1/2$ implies the ring part shrinks to the top of the lower $0$ branch and disappears. Conversely, $\left| c_{0} \right|^2$ decreases to $0$ and $\left| c_{-1} \right|^2$ increases to $1/2$ which means the ring part merges into the upper $-1$ branch [e.g., the case for $\omega=1.25$ in Fig.~\ref{fig:2}(a)]. Applying these critical conditions to Eqs.~(\ref{eq:hh10}) and (\ref{eq:hh11}), we obtain the frequency window of the existence of the ring structure as
\begin{align}
  -\frac{g}{2} + v \le \omega \le \frac{g}{2} + v\,.\label{eq:window}
\end{align}

In Fig.~\ref{fig:4}(c), we show the frequency window for the existence of the ring as a function of the interaction strength. We find the analytical result of the frequency window given by Eq.~(\ref{eq:window}) (with a blue dashed line) agrees well with the numerical results (red crosses) for small interaction strength ($g \lesssim 0.5$). For larger interaction strength, on the one hand, the numerical result gradually deviates from the analytical prediction (\ref{eq:window}) for the upper bound of the frequency window, since the increasing $g$ magnifies the effect of the other Fourier components which have been neglected in this approximation. On the other hand, the numerical result rapidly deviates from the analytical prediction for the lower bound of the frequency window. In this case, as $\omega$ decreases (i.e., the quasienergy space is narrower), more hybridizations occur between various branches. The mixing of more Fourier components can no longer be ignored, and hence the approximation made in Eqs.~(\ref{eq:hh10}) and (\ref{eq:hh11}) becomes invalid.

\subsection{``Multiple-ring" structures} 
\begin{figure}[th] 
  \centering \includegraphics[scale=0.6]{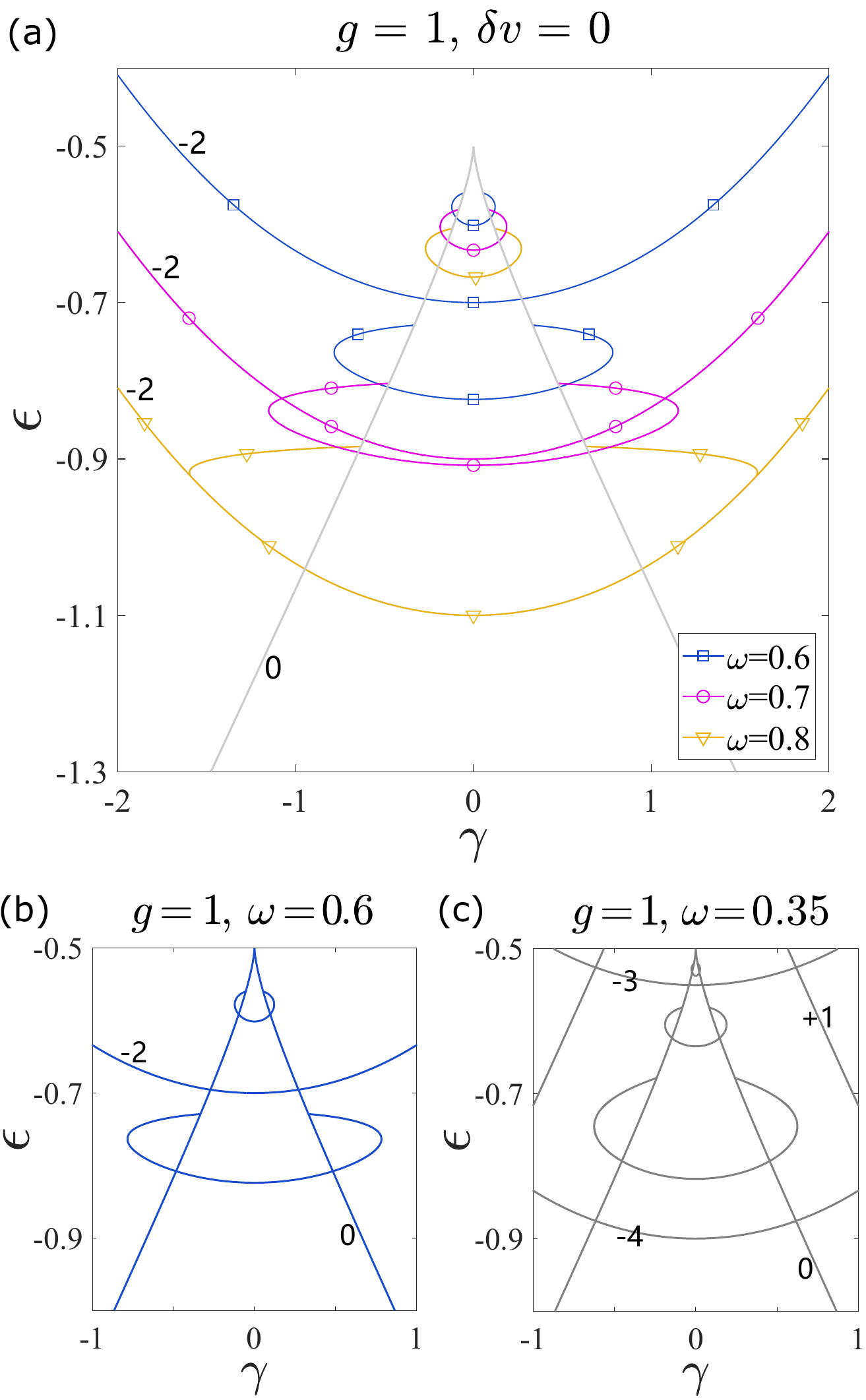} \caption{``Multiple-ring" structures. Here we take $g=1$ and employ the simplified analysis with $\delta v=0$. (a) Emergence of the ``double-ring" structure with decreasing $\omega$: the second ring appears when $\omega \lesssim 0.7$. The gray line shows the lower $0$ branch, which does not move by changing $\omega$. (b) The double-ring structure for $\omega=0.6$ [same plot as the one for $\omega=0.6$ in (a)]. (c) An example of the ``triple-ring" structure obtained for a smaller frequency $\omega=0.35$.}
	\label{fig:5}
\end{figure}

Next we consider the case of the low-frequency regime $\omega < 1$. Surprisingly, we find the appearance of a multiple-ring structure below the top of the lower $0$ branch; see, e.g., Figs.~\ref{fig:5}(b) and \ref{fig:5}(c) for a double- and triple-ring structure at frequencies $\omega=0.6$ and $0.35$, respectively.  

We show the evolution of the double-ring structure by decreasing $\omega$ in Fig.~\ref{fig:5}(a), which results in the emergence of the second ring. Similar to what has been seen in Fig.~\ref{fig:2}(a), the appearance of the additional ring structure is closely related to the shrinking triangular structures on the two sides about $\gamma=0$. For comparison, we take the same parameter values as in Fig.~\ref{fig:2}(b) except for employing the simplified analysis with $\delta v=0$. We see that while the first ring still remains for smaller $\omega$ [see the rings around $-0.7 < \epsilon < -0.5$ in Fig.~\ref{fig:5}(a)], the $-2$ branch moves upward as the frequency $\omega$ decreases (since the Floquet ``Brillouin" zone is continuously shrinking), and as a consequence the triangular structures from the two sides get shifted towards $\gamma=0$. In this process, the triangular structures continue to shrink until they give rise to a new independent branch enveloping the original branch at the critical frequency value ($\omega \approx 0.7$, in magenta); the new branch appears as the second ring structure. Similar behavior replicates itself at the next lower critical frequency which gives rise to triple- and multiple-ring structures.  

Following the same analysis in Sec.~\ref{sec.3b}, we find that the solutions of modes $a$ and $b$ in the second ring also have four dominant components, $c_{-2}$, $c_{0}$, $d_{-2}$, and $d_{0}$, which satisfy the following relation at $\gamma=0$:
\begin{align}
c_{-2}=d_{-2} \qquad \mathrm{and} \qquad c_{0}=-d_{0}.
\label{eq:hh13}
\end{align}  
Then we can get similar relations to Eqs.~(\ref{eq:hh10}) and (\ref{eq:hh11}):
\begin{align}
g\left| c_{-2} \right|^2  + \epsilon + \frac{v}{2} &=0,
\label{eq:hh14}\\
g\left| c_{0} \right|^2 + 2\omega + \epsilon  - \frac{v}{2} &=0,
\label{eq:hh15}
\end{align}
which give the parameter window of the existence of the double-ring structure.

As an example, we discuss the case shown in Fig.~\ref{fig:5}(a) ($g=1$). As $\left| c_{0} \right|^2 $ decreases to $ 0 $ and $\left| c_{-2} \right|^2 $ increases to $ 1/2 $, the upper $-2$ branch begins to touch the second ring at $\gamma=0$. For $g=1$, Eqs.~(\ref{eq:hh14}) and (\ref{eq:hh15}) give the critical values of $\omega=0.75$ and $\epsilon=1$ for the vanishing of the double-ring structure. This is in reasonable agreement with the numerical results in Fig.~\ref{fig:5}(a) at the critical case $\omega\approx 0.7$. Above the critical value of $\omega$, Eq.~(\ref{eq:hh14}) fails while Eq.~(\ref{eq:hh15}) still correctly describes the movement of the upper $-2$ branch at $\gamma=0$. For example, the quasienergy $\epsilon$ at $\gamma=0$ for the upper $-2$ branch at $\omega=0.8$ in Fig.~\ref{fig:5}(a) and the upper $-2$ branch in Fig.~\ref{fig:2}(b) are well described by Eq.~(\ref{eq:hh15}) with $ \left| c_{0} \right|^2=0$: $\epsilon = - 2\omega + (v/2)$.

\section{Adiabatic evolution}
In this section, we study the time evolution of the system under the periodic driving. It is well known that the system without time-periodic modulation follows the quantum adiabatic theorem: if the change of time-dependent parameter(s) in the Hamiltonian is sufficiently slow, the system initially prepared in an eigenstate of the Hamiltonian remains in its instantaneous eigenstate (e.g., Ref. \cite{Messiah_textbook}). The question arises whether the system under a time-periodic driving can trace the instantaneous Floquet eigenstate when other system parameter(s) are varied sufficiently slowly \cite{Weinberg17}.

\begin{figure}[h!] 
	\centering \includegraphics[scale=0.72]{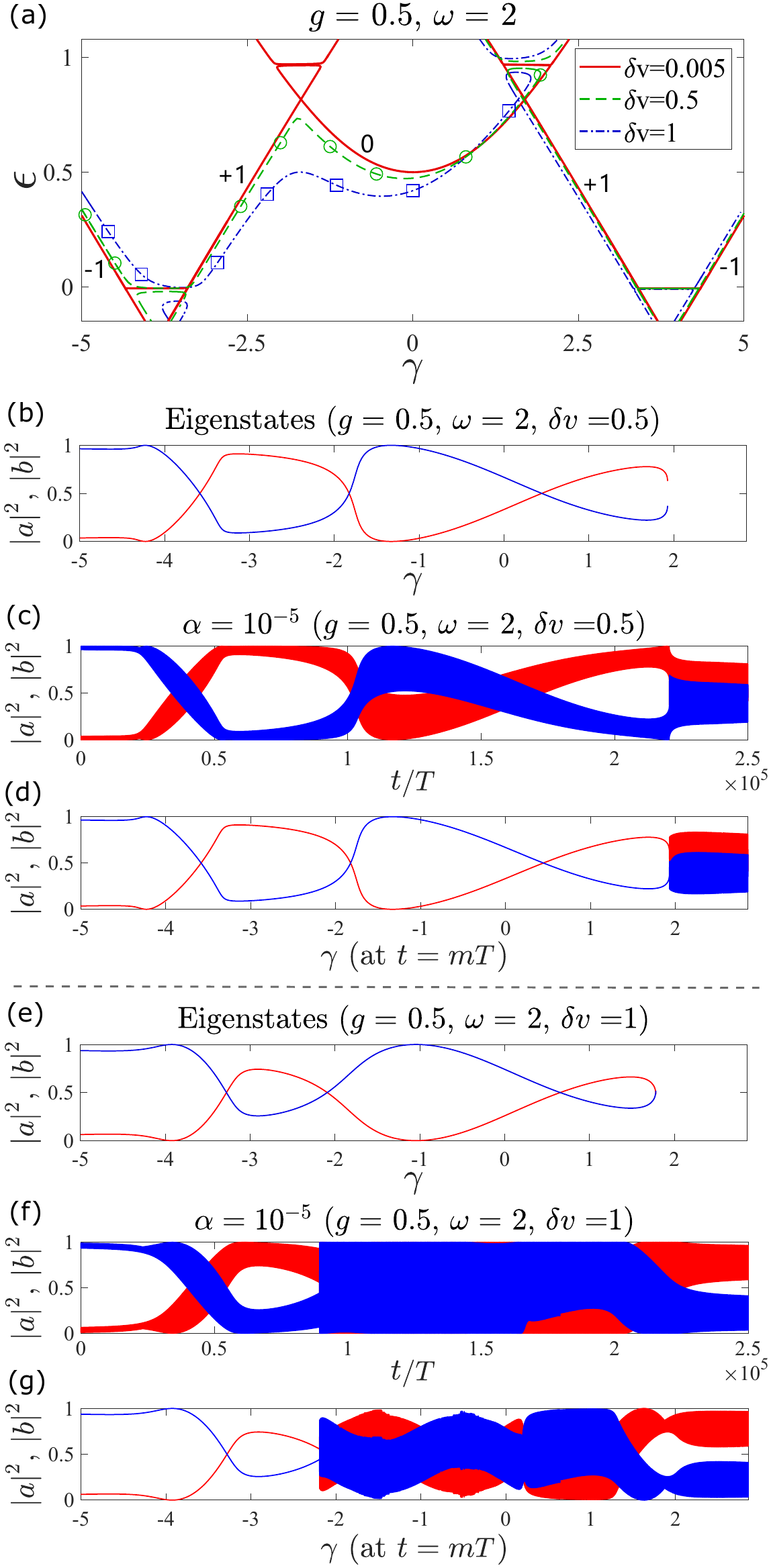} \caption{Approximate adiabatic evolution and the breakdown of the adiabaticity. (a) The quasienergy bands (marked by circles and squares, respectively) which we choose to perform the adiabatic evolution. (b) Populations of modes $a$ (in red [light gray]) and $b$ (in blue [dark gray]) of the Floquet eigenstates on the band marked by circles in (a). (c) Populations of modes $a$ and $b$ in the time evolution of the system. Here we start from a Floquet eigenstate (with a small perturbation) at $\gamma =-5$ on the same band as (b), and take the sweeping rate at $\alpha=10^{-5}$. (d) Same as (c) but for stroboscopic instances of time at $t=mT$ $(m \in \mathbb{N})$. (e)--(g) are the same as (b)--(d), respectively, but for another quasienergy band marked by squares in (a).}	
	\label{fig:6}
\end{figure} 

First, for comparison, we plot the instantaneous populations $|a(t=0)|^2$ and $|b(t=0)|^2$ of the two modes of the Floquet eigenstates as a function of $\gamma$ in Fig.~\ref{fig:6}(b), which correspond to the quasienergy band marked by circles in Fig.~\ref{fig:6}(a). We can see that the populations of modes $a$ and $b$ are sometimes inverted according to the band structure due to the hybridization of downward and upward Floquet branches.

For demonstration, we sweep the level spacing $\gamma$ in the following protocol:
\begin{align}
\gamma =  \gamma_{0} + \alpha\, t,
\label{eq:hh16}
\end{align}  
where $\gamma_{0}$ is the starting point of $\gamma$ and $\alpha$ is the sweeping rate. We start from the Floquet eigenstate with a small perturbation at $\gamma_{0}=-5$ in the quasienergy band marked by circles in Fig.~\ref{fig:6}(a), and sweep $\gamma$ at the rate $\alpha=10^{-5}$. The perturbation added to the initial state is $0.01\%$ in the populations of modes $a$ and $b$, i.e., $|a|^2 \rightarrow |a|^2 + 10^{-4}$ and $|b|^2 \rightarrow |b|^2 - 10^{-4}$. The time evolution of the populations of modes $a$ and $b$ is shown in Fig.~\ref{fig:6}(c), where a rapid oscillation of the populations can be observed. Figure~\ref{fig:6}(d) shows the populations of modes $a$ and $b$ at the moments of each integer period, i.e., at $t=mT$ $(m \in \mathbb{N})$, and its horizontal axis shows the corresponding value of $\gamma$.

Comparing Figs.~\ref{fig:6}(b) and \ref{fig:6}(d), we can see that the state of the system goes back to the instantaneous Floquet eigenstate after each period at $t=mT$. In other words, the system stroboscopically follows the instantaneous Floquet eigenstate, which demonstrates the adiabatic evolution in a stroboscopic manner. When $\gamma$ arrives at $\gamma \approx 1.93$, which corresponds to a terminal point of the loop in the band marked by circles [see Fig.~\ref{fig:6}(a)], the diabatic transition occurs since there is no state which can be adiabatically connected and the system shows a ``chaotic'' behavior.

Similarly, Figs.~\ref{fig:6}(e)--\ref{fig:6}(g) correspond to the case for the quasienergy band marked by squares in Fig.~\ref{fig:6}(a). However, Figs.~\ref{fig:6}(f) and \ref{fig:6}(g) show a ``chaotic'' oscillation of the populations even before the terminal point of the loop in this band. When reaching $\gamma \approx -2.18$, which is far from the terminal point of the loop, the Floquet eigenstate becomes dynamically unstable. Here, being dynamically unstable means that perturbations from the initial Floquet eigenstate expotentially grow in time so that the system cannot stay in the initial Floquet eigenstate even if the perturbations are infinitesimal but nonzero. The chaotic oscillation of the populations starting around $\gamma \approx -2.18$ would manifest itself as chaos in a Poincar\'e map analysis at this value of $\gamma$, which is also an indication of the dynamical instability (e.g., Refs.~\cite{Holthaus01a, equilibrium state and chaos}). Dynamically unstable states appear in the middle of a band unlike the time-independent nonlinear two-mode system \cite{Biao Wu NLZ, Biao Wu and Jie liu} due to the hybridization of Floquet branches. The unexpected emergence of dynamically unstable states in the middle of a band leads to the breakdown of the stroboscopic adiabatic theorem.

\section{Conclusion}
We have studied the Floquet eigenspectrum of a nonlinear two-mode system under a time-periodic driving in the off-diagonal terms of the Hamiltonian. We have found the triangular structures [see in Figs.~\ref{fig:1}(b) and \ref{fig:1}(c)] in the Floquet eigenspectra, which result from the combined effects of nonlinearity, the coupling of two Floquet branches, and the gap opening by the driving. Moreover, we have discovered completely different types of the exotic structures in the Floquet eigenspectra: ring, double-ring, and even multiple-ring structures [see Figs.~\ref{fig:2}(b), \ref{fig:5}(b), and \ref{fig:5}(c)], which result from the combination of nonlinearity and the hybridization of Floquet branches. In fact, such a combined effect may bring unexpected phenomena in the nonlinear driven system to which people have yet to pay much attention. Furthermore, we have clarified the mechanism of the emergence of these exotic structures of the Floquet eigenspectra and have provided an analytical prediction of the parameter window of their existence. In addition, we have demonstrated that the system under a time-periodic driving in principle follows the quantum adiabatic theorem stroboscopically. However, the stroboscopic adiabatic theorem can break down in the time-periodically driven nonlinear system due to the emergence of dynamically unstable states in the middle of the quasienergy bands.

In closing the paper, we shall discuss the experimental feasibility of our predictions. Taking the experiments on a BEC in a double-well potential done by the group at the University of Heidelberg \cite{Oberthaler08, Giovanazzi08njp}, we estimate the possible values of the parameters $v$, $g$, $\delta v$, and $\omega$ using their setup. First, in their paper by Giovanazzi \textit{et al.} \cite{Giovanazzi08njp}, the parameters $E_{J}/N$ and $E_{C}N/8$ correspond to our parameters $v/2$ and $g/2$, respectively. In their system, the total number of particles is $N=200$, $E_{C}/\hbar \approx 4.4\, \mathrm{Hz}$, and $E_{J}/\hbar \approx (0\, \textup{--}\, 3.7)\, \mathrm{kHz}$. Therefore, the corresponding values of $g$ and $v$ are $g/\hbar \approx 220\, \mathrm{Hz}$ and $v/\hbar \approx (0\, \textup{--}\, 37)\, \mathrm{Hz}$, respectively, and the range of the ratio $g/v$ is $g/v \approx (5.9\, \textup{--}\, \infty)$. Since $g/v \propto N^{2}$, we can easily go to the region of our interest, $g/v \sim 1$, by decreasing $N$. Next, we estimate the modulation frequency $\omega$ and the amplitude $\delta v$ based on the sweep rate and the sweep amplitude of the barrier height of their double-well potential. In the experiment by Est\`eve \textit{et al.} \cite{Oberthaler08}, they can control the barrier height accurately at least by of order $10$ Hz. According to Fig.~1 of Ref.~\cite{Giovanazzi08njp}, one can see that the $10$ Hz difference in the barrier height results in a difference in $E_J$ of order $1$\%, which corresponds to a minimum value of $\delta v /v$ = $O(10^{-2})$ in our model. Furthermore, they can ramp the barrier height such that the coupling between the wells is almost zero. This indicates that the maximum amplitude of $\delta v$ in our model could be comparable to $v$. Therefore, the range of $\delta v/v$ is of order $10^{-2}$ -- $1$, which covers the region of our interest. Finally, we estimate the frequency range of $\omega$ to modulate the barrier height by $\pm 100$ Hz [In this case, $\delta v /v$ = $O(10^{-1})$, which is still in the range of our interest.]. Since the barrier height can be controlled at the rate of order $1\, \mathrm{Hz}\, \mathrm{ms}^{-1}$ (to $1\, \mathrm{kHz}\, \mathrm{ms}^{-1}$) in their experiment \cite{Oberthaler08}, the frequency range of $\omega$ to modulate the barrier height by $\pm 100$ Hz is of order $10 \times 2\pi$ Hz (to $10 \times 2\pi$ kHz). Accordingly, $\hbar \omega/v = O(1)$ [to $O(10^3)$], which is in the region of our interest.

\begin{acknowledgments}
This work was supported by the NSF of China (Grants No.~11674283, No.~11975199, and No.~11974308), by the Zhejiang Provincial Natural Science Foundation Key Project (Grant No.~LZ19A050001), and by the Fundamental Research Funds for the Central Universities (Grants No.~2017QNA3005 and 2018QNA3004). L.-K.~L. and G.~W. are supported by the Zhejiang University 100 Plan, and by the Thousand Young Talents Program of China.
\end{acknowledgments}


\begin{thebibliography}{99}
\bibitem{note:drive} Instead of a double-well potential, if we use, e.g., two hyperfine states as the two modes $a$ and $b$, the off-diagonal driving of $\delta v\, e^{\pm i\omega t}$ can be realized by a time-periodic modulation of the Rabi coupling laser.
\bibitem{note:oscillation} Actually, there is an oscillation with time in the populations $|a(t)|^2$ and $|b(t)|^2$ and the relative phase between modes $a$ and $b$, due to the existence of the time-periodic driving. The oscillation has the same period $T$ as the driving, and the oscillation amplitude can be negligible provided the driving strength $\delta v$ is sufficiently small.
\bibitem{note:ring} The ring part consists of two curves as can be seen in the cases for $\omega=1$, $1.1$, and $1.2$ in Fig.~\ref{fig:2}(a). These two states become degenerate in the limit of $\delta v = 0$.
\bibitem{note:gap} In the on-resonance case ($\omega=1$) in Figs.~\ref{fig:2}(a) and \ref{fig:2}(b), there is actually a very tiny gap between the upper $-1$ branch and the lower $0$ branch, which is not discernible in the figures, due to a small but nonzero value of $\delta v$. In the limit of $\delta v=0$, the tiny gap vanishes.
\bibitem{note:cusp} More precisely, for a small but nonzero $\delta v$ as in the case of Fig.~\ref{fig:2}(b), the spectrum shows a cusp at $\gamma \approx 0$ when $g$ is slightly larger than $1$.
\end{thebibliography}
\end{document}